\documentclass[aps,pra,twocolumn,groupedaddress,showpacs]{revtex4}
\usepackage{graphicx}
\usepackage{dcolumn}
\usepackage{bm}
\begin{document}

\title{
Incompleteness and Limit of  Quantum Key Distribution Theory }

\author{Osamu Hirota}
\email{hirota@lab.tamagawa.ac.jp}
\affiliation{
 Quantum ICT Research Institute, Tamagawa University, \\
6-1-1, Tamagawa Gakuen, Machida 194-8610 Tokyo, JAPAN
}%

\date{\today}

\begin{abstract}
It is claimed in the many papers that a trace distance ($d$)  guarantees  the universal composition security in quantum key distribution (QKD). In this introduction paper, at first, it is explicitly explained  what is the main misconception  in the claim of the unconditional security for QKD theory. In general terms, the cause of the misunderstanding on the security claim  is the Lemma  in the paper of Renner. It suggests that  the generation of the perfect random key is assured by the probability $ (1-d)$,  and its failure probability is $d$.  Thus, it concludes  that the generated key provides the perfect random key sequence when the protocol suceeds. So the QKD provides perfect secrecy to the one time pad. This is the reason for the composition claim.
However, the quantity of the trace distance (or variational distance) is not the probability for such an event. If $d $ is not small enough, always the generated key sequence is not uniform. Now one needs the reconstruction of the evaluation of the trace distance if one wants to use it. One should first go back to the indistinguishability theory in the computational complexity based, and to clarify the meaning of the value of the variational distance. In addition, the same analysis for the information theoretic case is necessary. The recent serial papers by H.P.Yuen have given the answer on such questions.
In this paper, we show more concise description of Yuen's theory, and  clarify that the recent upper bound theories for the trace distance by Tomamichel et al and Hayashi et al are constructed based on the reasoning of Renner and it is unsuitable as the analysis for information theoretic security. Finally, we introduce a  macroscopic quantum communication with different principle to replace Q-bit QKD. 
\end{abstract}

\pacs{
}
                             
\keywords{Quantum key distribution,  Unconditional security, Trace distance, Key estimation attack}
\maketitle

\section{Introduction}

Quantum information science holds enormous promise for entirely new kinds of computing and communications, including important problems that are intractable using conventional digital technology. The most expected field is quantum cryptography.
But realizing that promise will depend on theoretical guarantee of the security and the ability to transfer an extremely fragile quantum condition. Here we would like to point out 
that, in general,  scientists are not familiar with practical applications in the real world.
QKD is an example of the stern realities of the field.
 
Despite enormous progress in theoretical quantum key distribution, many theory groups are still discussing the security proof for QKD [1,2]. 
Recently, M.Tomamichel  et al announced that in any practical implementation, the generated key length is limited by the available resources, and the present security proofs are not established rigorously in such a situation [3].
It means that "asymptotic resource assumption" cannot be met by practical realization.
Such a statement is welcome to acomplish the development of the real secure communication technology.

However, without the review of the incompleteness of the theory, it is repeatedly and persistently claimed that a specific trace distance criterion would guarantee universal composition security in quantum key distribution.
So far several criticism on such theory of QKD have been presented [4,5]. Unfortunately, almost all the theory group on QKD ignored these criticism. This is disagreeable in the development of science and technology.
Researchers are obliged to clarify "what is going on" in the discussion of the scientific theory. 
At present, there is no review paper on such a dispute. The purpose of this paper is to introduce a story of the argument on the recent theory of QKD and the criticism against them. In the section 2, we introduce the Shannon theory on the cryptography to confirm the basis of the concept of the information theoretic security. In the section 3, we introduce the fundamental concept of the current security theory of QKD by R.Renner [1].
In the section 4, we provide an evidence on which there is no theoretical proof of the 
unconditional security for any QKD, despite that many theoretical papers claimed the perfect proof of the unconditional security based on the Renner's concept. That is, we confirm the following statement:\\
\\
{\it One cannot ensure the information theoretic security of QKD only by the concept such as $\epsilon$-security. One needs to spell out the operational significance of the definition}.\\
\\
In the section 5, we give the outline of the Yuen's criticism. In the section 6, we explain the unsuitable performance of Q-bit quantum communication like QKD. In the section 7, we show why Q-bit quantum communication should be replaced by macroscopic quantum communications, and explain the macroscopic quantum effect of coherent state which is useful to apply it to secure communications.

\section{Theory of  cryptography  for Physicist and Mathematician}
In order to simplify the description, we treat a stream cipher that is the representative encryption method in the conventional cryptography. The transmitter and the receiver so called Alice and Bob share the pseudo random number generator(PRNG) with a short secret key ${\bf K}_S$. The information bit sequence ${\bf X}$ as the data is scrambled by XOR operation of the data bit and the output bit of the PRNG at the transmitter. The bit sequence of the output of the XOR operation is called ciphertext ${\bf Y}$. 
\begin{equation}
{\bf Y}={\bf X} \oplus {\bf K}_R
\end{equation}
Bob receives the ciphertext and he operates again XOR to the ciphertext by own output bit sequence from PRNG. Finally he can obtain the information bit sequence from such operations. 

Here we describe the fundamental symbols as follows: ${\bf K}_S$ is secret key, $|{\bf K}_S|$ is bit number of the secret key or key length,  ${\bf K}_R$ is the running key,  $|{\bf K}_R|$ is bit number of the running key or running key length.
 The role of the PRNG is to extend the secret key length, keeping a good randomness. 
For example, let as assume $|{\bf K}_S|$ =256 bits. The secret key is extended by appropriate PRNG to 
\begin{equation}
|{\bf K}_S| =256 \Rightarrow |{\bf K}_R| =2^{256}-1
\end{equation}
This corresponds to the running key. Thus, $2^{256}-1$ bits data can be encrypted only by 256 bits secret key. 

The secret key has to be generated from the physical random phenomena, because 256 bit should have uniform randomness. It corresponds that the estimation probability of the secret key before the encrypted communication is 
\begin{equation}
P_{suc}({\bf K}_S) =2^{-256} \sim 10^{-66}
\end{equation}
When the encryption system satisfies the above condition, the security analysis goes to the search of the decryption algorithm. The important parameters for such an analysis are the complexity of the structure of PRNG and the length of ciphertext that Eve uses to determine the secret key. In principle, it is decrypted when Eve gets the known plaintext (information data) of 256 bits and the corresponding 256 bits ciphertext by so called brute force attack. But its calculation number:$N$ that Eve has to try is 
\begin{equation}
N=2^{256} \sim 10^{66}
\end{equation}
Let us assume that Eve has the billion super computers. She may need more than million years to determine the secret key of 256 bits. Thus, it has no trouble in the real world application.
The main problem in such a mathematical cipher is that nobody can deny the existence of the algorithm to determine the secret key from the information of the structure of PRNG when Eve can get very long ciphertext or known plaintext. 
This is denoted such that the security level of the system does not rule out drastic breach of security.

Only one way to avoid such situation is to employ the one time pad stream cipher. 
It means that the running key sequence from PRNG is replaced by physical random sequence. 
Shannon provided the theory of the security for the one time pad as follows:\\
\\
{\bf Definition 1}(Shannon, 1949 [6])\\
The perfect secrecy or full information theoretic security is defined as follows:
\begin{equation}
P({\bf X}|{\bf Y})=P({\bf X}) \quad \forall {\bf Y}
\end{equation}
where $X$ is plaintext and $Y$ is ciphertext, respectively.
\\
\\
This is sometimes called unconditionally secure stream cipher. But this terminology is not adequate in the real world, because it is liable to cause misunderstanding.
\\
\\
{\bf Theorem 1}(Shannon, 1949 [6])\\
The perfect secrecy in the one time pad is attained when the following conditions are satisfied:
\begin{equation}
|{\bf K}_S| = |{\bf K}_R|\ge|{\bf X}|
\end{equation}
\begin{equation}
P_{suc}({\bf K}_S)=2^{-|{\bf K_S}|}=2^{-|{\bf K_R}|}
\end{equation}
\\
The first condition means that the secret key and the information bit must have the same length. 
The second condition means the perfect uniformity of the secret key. Also it corresponds to the estimation probability for the secret key before the secret communication.  
That is, the secret key must be generated from the physical complete random phenomena.

For example, when the key length is 10,000 bits, the estimation probability is as follows:
\begin{equation}
P_{suc}({\bf K}_S)=2^{-10,000} \sim 10^{-3,000}
 \end{equation}

In the real application, several Tera bits random sequence generated from the physical phenomena is stored in the Hard Disk and it is carried by the air plane et al. Thus, its system may provide the perfect uniformity and the potential of the perfect secrecy.
There is a big question.  Can one realize the key distribution by communications which can provides the perfect secrecy? A quantum key distribution QKD was suggested in order to give a solution for such a realization problem.
\\
\\
{\it The problem is that whether such QKD can provide the perfect uniform key sequence or not}.

\section{Formulation of QKD theory by R.Renner}

There is a long story on the security analysis of the quantum key distribution, and the theory has very complicated structure. But here we give concise description. 

\subsection{Definition of Security}

The claim that the generated key sequence provides the perfect secrecy is made on behalf of the trace distance criterion:$d$ as follows [1]:
\\
\\
{\bf Definition 2} \\
The trace distance is defined as follows:
\begin{eqnarray}
d \equiv \frac{1}{2}||\rho_{KE}-\rho_U \otimes \rho_E||_1
\end{eqnarray}
When 
\begin{equation}
d \le \epsilon
\end{equation}
the generated key is called "$\epsilon$-security". Some times, it is called unconditional security, if $\epsilon$ is small enough.\\
\\
Renner and his group claim that the generation of the perfect random key is given by the probability $(1-d)$,  and it failure probability is $d$. If one has the upper bound of $d$ as $\epsilon$, the probabilities for the success and failure are (1- $\epsilon$) and $\epsilon$, respectively.
Thus, they claim that the generated key sequence is always true uniform random bit sequence whenever the protocol well succeeds. This is justified by the following reasoning.

\subsection{Reasoning of the Interpretation}
The community of QKD employs the following variational distance formulation to give the reasoning. 
In the classical cryptography, the security is evaluated by the variational distance. 
\begin{equation}
\delta(P,Q)=\frac{1}{2}\sum_{x\in X} |P(x)-Q(x)|\le \epsilon
\end{equation}
Such a theory is called indistinguishability theory. 
Therein, the attacker's ability is limited to the polynomial computational power or resource. 
Under this assumption, one evaluates how much precisely the attacker can distinguish two random systems. The upper bound is related with the algorithm. 

On the other hand, in the case of information theoretic case, the attacker has unbounded computational resource. So under such a case, the upper bound of the variational distance must be evaluated. Renner first discussed the classical case and gave the following Lemma.
\\
\\
{\bf (Lemma of Renner [1])} \\
Let $P$ and $Q$ be two probability distributions. Then there exists a joint probability distribution $P_{XX'}$, such that $P_{X}=P$, $P_{X'}=Q$, and 
\begin{equation}
Pr[x \ne x']=\delta(P,Q), \quad P_{XX'} \rightarrow (x,x')
\end{equation}
\\
Renner deduced the following statement from the above.
\\
\\
{\bf (Statement)} \\
The variational distance between $P$ and $Q$ can be interpreted as the probability that two random experiments described by $P$ and $Q$ , respectively, are different. 
It can apply to the quantum case.
\\
\\
This leads a great misconception in the security analysis for QKD.

\section{What is wrong in QKD theory?}
\subsection{Fundamental Concept}
The statement in the above section is derived from the Lemma of Renner that asserts the existence of a joint distribution which gives marginal distributions $P$ and $Q$ and for which the results of the two random experiments differ with just probabilities  $\delta(P,Q)$. 

H.P.Yuen has repeatedly claimed that this does not imply the Renner interpretation of $\delta $  which is the 
basis of his $d$ interpretation [5,8,9]. The $\epsilon$ is not an event probability though it may be the difference of two event probabilities, and the variational distance is not the probability that information is leaked. More important fact to deny the interpretation of Renner is as follows [4,5]:
\\
\\
{\bf Lemma 1} \\
When $d$ is not zero, the generated key sequence is not uniform. 
\\
\\
That is, one cannot claim the uniformness of the generated key sequence  based on the Renner's concept. Thus, the interpretation of $\epsilon$ employed by the whole community of QKD such as "failure or success" probability of the protocol is wrong.
This fact gives very serious effect to the theory of the upper bound for the trace distance. We show an example in the next section.

\subsection{Misconception of Upper Bound Theory of the Trace Distance}

Here we describe the concrete misconception on the trace distance and its upper bound.
The trace distance is a measure for the nearness between density operators or probability distributions. It may be bounded by several theoretical functions. For example, $d$ is clearly bounded by Holevo quantity as follows:
\\
\\
{\bf Theorem 2} [7]\\
The trace distance is bounded by the Holevo quantity  $\chi$ as follows:
\begin{equation}
 2d^2 \le  \chi 
\end{equation}
where
\begin{equation}
\chi =S(\rho_E) - <S(\rho^K_E)>_K
\end{equation}
Furthermore, there are many options of the upper bound.
One can accept such a formulation. 
But the main problem is to evaluate the quantitative feature of the upper bound.
However, the unsuitable method has been employed. 
\\
\\

Here we would like to clarify what is the problem.
They misunderstood the conception of the cryptography based on the indistinguishability theory as denoted in the section 3, and under such concept they invented several methods to give the upper bound. 
 Renner gave a quantum generalization of the classical Leftover Hash Lemma, and 
it may give a theoretical justification of the role of the privacy amplification.  Then his group made a theory of smooth mini entropy and so on to give a numerical property of the upper bound of the trace distance.
That is, they confirmed that the trace distance provides a kind of failure probability of the protocol. Hence, if they can estimate the above parameters related with the failure probability of the protocol, they can think that it gives the upper bound of the trace distance.
\\
\\

In 2009, Cai and Scarani [10] tried to calculate the the upper bound based on smooth mini entropy and privacy amplification, and Scarani simplified its explanation [11].
They explicitly described in their paper [10] such that the $\epsilon$ has the operational meaning: it represents the maximum probability of failure that is tolerated on the key extraction protocol.
The value of $\epsilon$ (the upper bound of $d$) is assumed without any reason.
Then, they constructed the upper bound theory based on the concept of the failure probability and related parameters.  
Thus,  we can clearly see that these come from the fact that they employ Renner's concept.

Yuen showed that the concept of the estimation probability on key is an essential to ensure the security in his serial papers [4,5,8,9]. Then, in 2011, Tomamichel et al [3]  gave the statement as follows:
\\
\\
({\bf Statement}) \\
Despite enormous progress in theoretical QKD, the security of the present scheme is not  established rigorously.
\\
\\
So they changed own theory such that the upper bound of the trace distance is given by a function $\Delta$ of the smooth mini entropy. 
\begin{equation}
\Delta =\min_{\epsilon '} \frac{1}{2} \sqrt {2^{l-H^{\epsilon '}_{min}(K|Y)}} +\epsilon '
\end{equation}
where the smooth mini entropy has an operational meaning from the quantum leftover Hash lemma 
such that it is related with Eve's estimation probability of the key sequence. Despite their effort, their function $\Delta$ and the trace distance $d$ 
are compulsorily connected as follows:
\begin{eqnarray}
&& d \le \Delta \\
&&(1-p_{abort})\Delta \le \epsilon
\end{eqnarray}
where $p_{abort}$ is the probability for aborting the protocol.
Then $\epsilon$ is given as  "given security level" without any reason. They assume  $\epsilon=10^{-10}$ as the typical value.
If it is a failure probability, it is sufficient value.
Consequently, it seems that they still keep in mind the concept of the failure probability from the above equations. Thus, they go back to a concept of "failure".
Since the trace distance does not mean the failure, they cannot ensure the security.
\\
\\

Contrarily Hayashi-Tsurumaru keep the original way, and 
give the following direct method to evaluate  the upper bound of the trace distance [12].
They define that $\epsilon$ is the parameter related with probability of the error in the estimation of the phase error rate from sample bits. 
In their paper,  $\epsilon=\sqrt \epsilon_{HS}$ is regarded as the upper confidence limit of the phase error as follows:
\begin{equation}
Pr\{c|{\hat p}_{shift}(c)\ge p_{shift}(k,c)\} >1-\epsilon_{HS} \quad \forall k
\end{equation}
where $c$ is error bits in sample, $k$ is error bits in the total bit, $p_{shift}(k,c)$ is  phase error rate of shifted key, and 
${\hat p}_{shift}(c)$ is  estimation of phase error rate of the shifted key, respectively.
Then they introduce the statistical fluctuation to proceed the logic.
They employ the normal distribution function to treat the statistical fluctuation 
in the estimation of the phase error rate as follows.
\begin{eqnarray}
\Phi (s(\epsilon _{HT}))&=&\int_s^\infty \frac{1}{\sqrt 2} exp(-x^2/2)dx \\
\epsilon _{HS}&=&\Phi (s(\epsilon _{HT}))
\end{eqnarray}
When $\epsilon _{HS}=\epsilon ^2$ is given, $s(\epsilon _{HT})$ is derived.

Finally they show that the bound is given by the decoding error probability in the phase error correction code. Of course they give the relation between estimation error of phase error rate and the decoding error based on the estimation error.
Consequently, the upper bound of the trace distance is given by the decoding error probability $P_{phase}$ of phase error correction.
\begin{equation}
d \le  \sqrt 2 \sqrt P_{phase}
\end{equation}
\\
Thus, their upper bound theory seeks parameters of failure or abort of the protocol.
So, their bound is effective when and only when the interpretation of the trace distance is related with the Renner's concept.
In fact, their numerical example clearly shows their concept. 
That is, also they just assume 
\begin{equation}
\epsilon _{HS}=4\times 10^{-26} \rightarrow s(\epsilon _{HS})=10.5
\end{equation}
Consequently the upper bound of the trace distance is 
\begin{equation}
d \le 10^{-10}
\end{equation}
Again if this means the failure probability, it make a sense, but it is too large  for the uniformness evaluation as shown in the next section.
Thus, they keep the concept of the failure in mind when they construct the upper bound theory for the trace distance.

One can invent many theories of the upper bound. Even if these can bound the trace distance, they cannot guarantee the security because of their misconception. 
That is, we emphasize that one cannot guarantee the information theoretic security of QKD only by 
\begin{equation}
\epsilon-security :\quad d \le \epsilon 
\end{equation}
\\
{\bf Remark} \\
$\epsilon$-security provides the relative evaluation for protocl, but it does not provide the absolute evaluation as the cryptographic matter.\\

The community of QKD avoids the discussion of the real issue of the cryptography, and 
they insist as if the $\epsilon$-security provides unconditional security or information theoretic security in the real application.  

\section{Key estimation attack for QKD}
\subsection{Basis}
In the previous section, we have explained the incompleteness of the security theory of the QKD. In order to improve the situation, here we introduce the reformulation of the trace distance which unifies the indistinguishability and the Shannon's definition of the perfect secrecy.
In the unified theory by H.P.Yuen, the most important attack is the estimation attack against the non uniform key.
According to the lemma 1, the generated key sequence in QKD is, in general, not uniform. 
So Eve will try to estimate the generated key sequence based on her any quantum measurement to 
photon streams that carry the raw key. 
This key estimation attack for QKD is formulated by the estimation probability on the key generation process. 
Let ${\bf Y}_E$ be the random bit sequence as the result of her quantum measurements. 
The success probability of estimation of the key sequence is denoted
as $P({\bf K}_G | {\bf Y}_E)$ based on quantum detection theory.
\\
\\
{\bf Definition 3} \\
If the success probability of estimation of the key sequence ${\bf K}_G$ is 
\begin{equation}
P({\bf K}_G | {\bf Y}_E) \sim 2^{-{|\bf K}_G|} \quad \forall  {\bf Y}_E
\end{equation}
then the system is information theoretically secure and it has  composablilty. \\
\\
In the above definition, one needs the perfect uniform distribution of key sequence in QKD protocol to attain the secrecy. This definition is natural according to the Shannon's concept on the information theoretic security, and the generated key provides the perfect secure one time pad.

As the next step, one needs to evaluate the upper bound of the success probability to guarantee the security. Fortunately, 
the success probability of key estimation can be related with the trace distance.
\\
\\
{\bf Theorem 3} [8,9] \\
Let us assume $d \le \epsilon$. 
The upper bound of the  averaged success probability is given by 
\begin{equation}
<P({\bf K}_G | {\bf Y}_E)> \le \epsilon  + 2^{-|{\bf K}_G|}
\end{equation}
\\
The bound Eq(26) can be achieved after (26) with equality.
The meaning is as follows:  The upper bound of the success probability is evaluated by the trace distance or its upper bound.
In general, in order to give the bounds for $P({\bf K}_G | {\bf Y}_E)$, one needs full information on error correction code and privacy amplification scheme.

\subsection{Example}

Although the upper bound theories by Renner group and Hayashi group are not appropriate to evaluate the trace distance, we here employ it, and let us see what is happen. 

In general, they give a bound of $d$ as the average over the privacy amplification. So we put it as follows:
\begin{equation}
d \equiv <d_{PA}> \le \epsilon =10^{-10}
\end{equation}
From the corollary, 
\begin{equation}
<P({\bf K}_G | {\bf Y}_E)>  \le 10^{-10} 
\end{equation}
When the length of the generated key is $|{\bf K}_G|=10^4$, the security requirement of the key estimation probability is 
$10^{-3000}$ from Eq(25).  So $10^{-10}$ is excessively large and it does not work as the security guarantee. That is, 
\begin{equation}
10^{-3000}<< 10^{-10}
\end{equation}
In addition, one should apply the Markov inequality to obtain an individual privacy amplification guarantee.
This converts the averaged one to an individual one.
\\
\\
{\bf Theorem 4} [8,9]\\
Let us assume $<d_{PA}> \le \epsilon$. 
From two times application of Markov inequality, one gets 
\begin{equation}
P({\bf K}_G | {\bf Y}_E) \le \epsilon^{1/3}  + 2^{-|{\bf K}_G|} 
\end{equation}
\\
\\
When we use the result Eq(27), we have at worst case as follows:
\begin{equation}
P({\bf K}_G | {\bf Y}_E) \le 10^{-3.3}
\end{equation}
\\
If one requires $10^{-3,000}$ in Eqs(19,20), clearly these do not work.
Thus, even if we consider any favorable treatment of the present QKD theory, it comes to grief as the security guarantee.

\subsection{Proper Security Evaluation of QKD}
The real guarantee of the security of QKD is to show that it is comparable to that of a uniform key. In the above section we explained that the upper bound theories of the trace distance by the current researchers do not provide the requirement of the unconditional security. If the upper bound of the trace distance is evaluated correctly, it may have a meaning in the security analysis. But the present theory only gives the value of Eq(32). We show such a situation in the Table-1.

\begin{table}[htbp]
  \begin{center}
    \begin{tabular}{|c|c|c|}
      \hline
               /  &Present QKD & Requirement \\ \hline
   Key estimation & $10 ^{-3.3}$& $10^{-3,000}$\\ \hline
           \end{tabular}
  \end{center}
  \caption{Key estimation probability evaluation for the present theory and its requirement for the unconditional security}
\end{table}

\section{Non efficacy of Q-bit quantum communication}

Recently, some program administrators of science and technology became aware of the non efficacy of Q-bit communication in the practical use, and they have proposed "macroscopic quantum communication". In this section, we introduce a reason.

\subsection{Poor Communication Performance}

Let us explain the mismatch between the communication performance of QKD and the real communications.
The optical communications based on bright coherent states routinely achieve unsecured communications rates exceeding $10^{10}$ bits per second (10 Gbit/sec) over distances exceeding 10,000 Km. For the data center communication, it provides $10^{11}$ bits per second (100 Gbit/sec)  by the wavelength division multiple system as shown in the table-II

In general, to realize such a performance, one needs the average photon number per pulse at the transmitter as follows: 
\begin{equation}
<n> =10^6 \quad (photon/pulse)
\end{equation}

The primary purpose of quantum communication was to devise a method for the protection of information carried by such a high speed communication with the above large signal energy. That is, 
a role of physical encryption is to protect such a high capacity optical communication based on physical phenomena. So the encryption scheme has to have also high speed performance.

\begin{table}[htbp]
  \begin{center}
    \begin{tabular}{|c|c|c|}
      \hline
               /  & Data speed & Distance \\ \hline
   Basic system & 10 Gbit/sec& 10,000 Km \\ \hline
     Data center & 100 Gbit/sec (WDM)& 1,000 Km \\ \hline
    \end{tabular}
  \end{center}
  \caption{Performance of Conventional Optical Communication, and Target of Secure  Quantum Communication without Relay Station.}
\end{table}

The communities of quantum information made a story that 
Q-bit quantum communications are, in principle, capable of providing a provably secure communications channel, and that communications protected by quantum security can typically only be attacked "in transit" and are not vulnerable to off-line attacks at some point in the future using newly developed techniques or computational resources.
A solution for the above purpose of the community was the quantum key distribution by Q-bit communication.
To protect the data, they employed one time pad based on the key sequence generated by QKD. 
This is called hybrid cipher.

Let us consider the performance of the present hybrid cipher consisting of QKD and one time pad. In QKD process, Q-bit signals such as single photon have proven extremely fragile in the face of loss and noise, effectively limiting the range of quantum communications to  $10^3$ bit per second (Kbit/sec) at a range of 100 Km. 
So the data encryption speed is the same as that of QKD. This corresponds to the communication in the Stone Age.

\begin{table}[htbp]
  \begin{center}
    \begin{tabular}{|c|c|c|}
      \hline
               /  & Data speed & Distance \\ \hline
   Basic system & 1 Kbit/sec & 100 Km \\ \hline
   WDM system & 10 Kbit/sec (WDM)& 100 Km \\ \hline
    \end{tabular}
  \end{center}
  \caption{The communication performance of QKD. There is the strong trade-off between speed and distance.}
\end{table}

Such a limit of the performance is verified by the fundamental physical law of communication 
which is called quantum communication theory.
Thus, there is no method to improve the efficiency of the communication if one employs Q-bit with such strong quantum effect as signals.  

The real world is asking the secure optical communication by means of physical encryption for communications as shown in the Tale II.  If one uses one time pad, the speed of QKD has to be the same as the data speed of the conventional optical communication. But, there is no possibility of realization of such a performance by QKD.

If the security technology for the low speed communication is a target, one does not need a new technology to protect the data, and it can be done by means of a hybrid cipher of one time pad and Hard Disk (random bit sequence)  carried by bike.
Thus, the role of quantum communication should be to 
provide the secure high capacity optical communication.

\subsection{Trick of Key Rate Theory}

If one wants to realize the real requirement, one has to realize QKD system over $10^9$ bit per sec (1 Gbit/sec) as the signal transmission speed, because one needs one time pad for 1 Gbit/sec. If one uses the generated key as the secret key for the conventional mathematical cipher such as AES (Advanced Encryption Standard), the security of the total system is the same as the mathematical cipher such as AES.

Any theoretical group knows such a problem, but nobody confesses such a defect of Q-bit communication. Their interest was devoted only to the bit per symbol rate of the key generation. 
However, in all papers, they ignore the energy loss effect to evaluate the key rate.
The rate is evaluated by the received photon sequence, not transmitted photon.
In the conventional communication, the almost all signals arrive at the receiver, so one can evaluate the rate by the received signal sequence.
In the Q-bit quantum communication, if the transmission distance is 100 Km, the energy loss is 20 dB, so already the rate becomes about 0.01. Although many papers claimed the Shannon limit for the coding et al, these do not make a sense in the practice. 

If one wants to provide the system applicable to the real world, the important unit is bit per second, not bit per symbol. 

\subsection{Quantum Repeater for QKD}
The devices that create Bell pairs over a distance are called "quantum repeater", building on the concept of teleportation. In general, the form of the repeater consists of the purification and swapping.
Bell pairs are consumed during the course of teleportation, purification, and entanglement swapping. Thus, the primary job of a repeater is to continually produce new ones.
There is no doubt that the concept of the quantum repeater is a great idea in the quantum information science. 
However, although one can enjoy the terminology of the information science in quantum physics, 
one cannot easily connect it with the information science for the real world.
The reason is the same as QKD itself. The real efficiency suffers the energy loss. 
Finally the speed of the repeater system becomes extremely slow. For example, for the transmission of 1,000 Km, when the present QKD and the repeater system are employed, the signal transmission speed is 
\begin{equation}
R=10^{-1} \quad (bit/sec)
\end{equation}
This means that one has to wait over 30 years to share $10^9$  bits (1 Gbit).

\subsection{No Security Guarantee of Real System}
As we discussed in the sections 4 and 5, the quantum key distribution protocol has no security guarantee even in the theoretical model at present. 
In addition, almost all experiments do not show the quantitative security parameter, despite that they claim unconditionally secure. This is unusual.

If they wants to quantify the security of the concrete protocol, they need to employ the estimation probability on key, providing the concrete error correction code and privacy amplification code. So, they have to abandon an indirect parameter such as $\epsilon$-security which is not accepted in the real application. It works only for mathematical interest.

Since the improvement of the communication performance and the security cannot be expected by the physical law of the communication, 
these technologies should be limited to a model for the physics experiment in universities to study the principle of quantum theory.

\section{Macroscopic quantum communication}
 In this section, we explain a research program of technology which can overcome the defect of Q-bit quantum communications.
\subsection{Basic Concept}

The present information security is algorithmic, and as a result, not provably secure. Crypto-systems of algorithmic security include pseudo random number generation and public key encryption. The security of these algorithmic techniques is based on the assumption that certain mathematical problems are effectively impossible to solve using contemporary computer resources and well-known attacks. However, this type of security is in-principle vulnerable to off-line attacks.
If the data speed is not so high as the present wireless communication, the present information security technology can protect the data or secret key against any enemy. 
However, in the current optical communication, the data speed is over 100 Gbit/sec, and 
any present scheme cannot encrypt such data with provable security. So we should inquire a new technology to cope with the above problem.
\\
\\

The physical direct encryption is a candidate for such a purpose. In order to accomplish such problems, one needs to develop macroscopic quantum communications. 
In 1990, we established the international conference on quantum communication, measurement and computing (so called QCMC) in order to discuss a potential of macroscopic quantum communications.

Our group is developing  innovative research in the area of macroscopic quantum communications (proposals which can combine the security of quantum communications with the distances/rates of macroscopic telecommunications). 
This research provides innovative approaches that enable great advances in secure quantum communications.

The primary goal of our program is to demonstrate that quantum communications can communicate  information data at sustainable rates of 1 Gbit/sec to 100 Gbps at distances of 1,000 Km. Our program also has goals as follows.\\
\\
${\bf (1)}$: To demonstrate that secure quantum communications can be extended to entirely new domains, such as sea and space. \\
\\
${\bf (2)}$: Since one time pad is meaningless in the modern communication, we extend quantum communications beyond 
key distribution to other practical, scalable quantum protocols. \\
\\
${\bf (3)}$: One should use purely classical means of encoding bright coherent states with encrypted information (e.g., using classical phase or amplitude modulation, or relying on pseudo-random algorithms), but its security performance must be protected by " ${\bf full}$ ${\bf quantum}$ ${\bf effect}$".
\\
\\
${\bf (4)}$: It is not allowed to extend the distance of communications using relay stations  in between the transmitter and the destination, except for intermediate optical amplifiers.\\
\\
We denote that utilizing entanglement is not planned, because of its extremely fragile in the face of loss and noise. 
It cannot realize the requirement from the real communication performance such as 10 Gbit/sec over 1000 Km, because of physical law of the communication.\\

Our programs include plans for a laboratory-scale testbed capable of conclusively demonstrating this scalability using the fiber cable system in our university. All elements necessary for continuous operation at the speed of data and distances above can be addressed, with prototypes of each key technology delivered to the testbed. Our plan produces prototypes and deliver them to a central testbed in the collaboration with the cable companies. 

\subsection{Macroscopic Quantum Effect}

The technologies which could be critical include: high-rate deterministic sources of single photons and entangled photons. High-rate single-photon detection is useless for the real communications, because the system must clearly demonstrate how each of the following rate/distance will be achieved:\\
\\
${\bf (1)}$:Direct data encryption over 10 Gbit/sec, not key distribution, not one time pad.\\
\\
${\bf (2)}$:Communication at 1$\sim$10 Gbit/sec over 1,000$\sim$10,000 Km without the stations.\\
\\
${\bf (3)}$:The use of the cryogenic cooling is not allowed.\\
\\
The  quantum communications including Q-bit are highly sensitive to loss and leads difficult situation on realization of the requirement. \\
Question arises: what quantum effect is useful to realize the above requirement ?.
\\
Let us consider the uncertainty principle as the most important quantum phenomena.
The ${\bf uncertainty}$ ${\bf principle}$ has two structures:\\
\\
(i) Kennard-Robertson indeterministic principle \\
(ii) Heisenberg uncertainty relation\\

The latter is used in the QKD to evaluate the disturbance by Eve. This is a typical example of the microscopic quantum effect. But the former does not imply the microscopic, it is also effective in the macroscopic region. 
A typical example is non orthogonal quantum states in optical modes like coherent state or squeezed state. 
The coherent state is the minimum uncertainty state for two conjugate observable like quadratic amplitude of light field that implies the Kennard-Robertson indeterministic principle [13]. 
Also it cannot be cloned with the fidelity 1 according to the ${\bf no cloning}$ ${\bf theorem}$ [14]. On the other hand, 
if one measures certain observable of light wave with coherent state, one suffers quantum noise effect. Consequently, one cannot discriminate information signals carried by coherent state without error according to the ${\bf state}$ ${\bf indistinguishability}$ ${\bf theorem}$ [15]. 
The above quantum effects are observed not only in the case of weak signal but also in the case of strong signal. For example, let us assume two coherent amplitudes $\alpha_1, \alpha_2$ of the coherent states, and $\delta =|\alpha_1- \alpha_2|<<1$, and $|\alpha_1| >>1, |\alpha_2| >>1$. 
The ultimate error in quantum detection is [15]
\begin{equation}
P_{error}=\frac{1}{2}[1-\sqrt {1-exp(-|\delta|^2})] \sim \frac{1}{2}
\end{equation}
Strong laser lights with very close amplitude cannot be distinguished by the quantum effect.
Also these cannot be cloned. If the attacker suffers such an effect in communication process, 
we can use this quantum principle for the secure communication [16] [17].
Thus coherent state is applicable to such concept, because  coherent state is the most robust against any decoherence in optical channels, and it has macroscopic quantum effect.

Here we can conclude that the key technology in quantum information science is to manipulate the macroscopic quantum effect that forces bad effect to the attacker, but allows the classical optical communication to the legitimate users. It is a kind of wire tap channel as follows:
\\
${\bf (1)}$:The channel of Alice - Bob = Conventional optical communication performance over 10 Gbit/sec.
\\
\\
${\bf (2)}$:The channel of Alice - Eve = Quantum communication performance based on several quantum theorems.\\

\begin{table}[htbp]
  \begin{center}
    \begin{tabular}{|c|c|c|}
      \hline
               /  & Feature & Latency \\ \hline
   Channel of Alice - Bob & Classical & No \\ \hline
    Channel of Alice - Eve & Quantum effect & /\\ \hline
    \end{tabular}
  \end{center}
  \caption{A Scheme of Macroscopic Secure Quantum Communication required by the Ministry of Defense}
\end{table}

A channel of Alice - Bob transmits the binary coherent states $|\alpha(1)>$ or $|\alpha(1)>$, but two signals are randomized by a mathematical method, and it can be decrypted by the secret key. A channel of  Alice - Eve, for example,  is described as follows:

Let us assume that the discrimination among quantum states at the output of the channel is described by POVM.
\begin{equation}
\Pi^E_j \ge 0, \quad \sum \Pi^E_j =I,
\end{equation}
where $I$ is an unit operator, $i$ and $j$ are $1,2,\dots M$.
Then a conditional probability for each trial of the measurement is given by 
\begin{equation}
P(j|i)=Tr \rho_i \Pi^E_j.
\end{equation}
where 
\begin{equation}
\rho_i =|\alpha_i><\alpha_i|
\end{equation}
By the mathematical randomization, the signal states become mixed state, but the probability is caused by the mathematical scheme.

The minimization problem of the average error probability based on the above equation is called quantum detection theory, which is a fundamental formalism in quantum information science. If the average error probability is given by
\begin{equation}
P_e =\min_{\Pi} \{1-\sum p_i Tr\rho_i \Pi^E_i\}\sim 1-\frac{1}{M}
\end{equation}
This performance is completely determined by full quantum nature.
And Eve cannot obtain any information. 
Although Eve can extend her measurement to the collective quantum decision making scheme, one can design to protect such attack.

Thus, the channel of Alice- Bob looks like a classical one which can provide high capacity communication such as 1 Gbit/sec to 100 Gbit/sec, but the channel of Alice - Eve who wants to get the information is blocked by the quantum effect. The summary is shown in the table-IV.

\subsection{Initial Shared Key Problem}
In physical cryptography, there is no possibility to share the key sequence without the initial shared key as shown in the table V.

\begin{table}[htbp]
  \begin{center}
    \begin{tabular}{|c|c|c|c|}
      \hline
               /  & Initial shared & Purpose & Security   \\ \hline
   BB84 & $|K_s|> 256 $ bits & Authentication & Intrusion detection  \\ \hline
   New  & $|K_s|$ = 256 bits & Seed of PRNG & Quantum noise hide  \\ \hline
    \end{tabular}
  \end{center}
  \caption{A role of the initial share key. In BB84, the security for the ket extension process has to be guaranteed. In a new one, the seed key and running key from PRNG have to be hidden by quantum effect. }
\end{table}

Here we can say that the glamorization of QKD faded when the proof of the necessity of initial shared key was given. That is, nobody can start the protocol for key distribution without the initial secret key, and it is a retreat of the function for the requirement from the real world.
If the initial shared key is unavoidable, it is valuable that  researchers of Quantum Information Science consider how to control the quantum effect by means of PRNG as the quantum symmetric key encryption.

\section{Conclusion}
We have introduced a theory that the claim of the strong security on quantum key distribution is incorrect. From the simple explanation, we believe that one could understand what is wrong in the theory of QKD. 
Although the trace distance plays an important role in the security analysis, it is not sufficient to guarantee the information theoretic security.
The role of the quantum key distribution is to provide the secret key sequence for the symmetric key cipher including one time pad. So the quantitative evaluation against the key estimation attack is necessary, according to the Shannon's original idea on information theoretic security.
Thus, the current theory of QKD does not work at all, and a development of  the evaluation theory of the key estimation based on quantum detection theory is necessary.

Even if the theoretical issue is solved, one has serious problem.
That is, Q-bit quantum communication to realize QKD is useless for the commercial system because the communication performance is excessively poor. To put it simply, the key distribution by Hard Disk may work well and it provides the unconditional security which is stronger than QKD.
Thus, we should develop macroscopic quantum communications for the real application of quantum information science to the secure communication.
This should be done urgently, otherwise quantum technology cannot contribute to the current high capacity network with 100 Gbit/sec data transmission which is promptly asking the ultimate security.

\section*{Acknowledgment}
This work was supported  by the special funding for the project on Macroscopic Quantum Communication of the Tamagawa University.
This paper is an extended version of Proceedings of SPIE conference on Quantum Communication and Quantum Imaging 2012, and a part of the presentation in The Ministry of Defense of Japan at Oct. 2011.

\section*{Appendix}

\subsection{Response to the paper}

We have several response on this paper as follows:\\
1. Science Daily: Quantum cryptography theory has a demonstrated security defect.\\
(http://www.sciencedaily.com/)\\
2. Help Net Security: Quantum cryptography theory has a proven security defect.\\
(http://www.net-security.org/secworld.php?id=13383)\\
3. Asia Pacific Security Magazine: Quantum cryptography theoru has a proven security defect.\\
(http://www.asiapacificsecuritymagazine.com/)\\
4. Global security magazine:Quantum cryptography theoru has a proven security defect.\\
(http://www.globalsecuritymag.com/)\\
5. Nanotechnology Now: Quantum cryptography theoru has a proven security defect.\\
(http://www.nanotech-now.com/news.$cgi?story_id$=45724).\\
6. Infosecurity magazine:Quantum cryptography theoru has a proven security defect.\\
(http://www.infosecurity-magazine.com/)\\
7.Quantum.com:Study proves Incompleteness and limit of security theory in quantum cryptography.
\\
(http://www.azoquantum.com/news.aspx?newsID=76)\\
8.Communications for automation: Defect found in quantum cryptography.\\
(http://www.isssource.com/)\\
9. US Department of Homeland Security: Daily Report August 8, 2012, p-17.\\
10. North DAKOTA Homeland Security:August 8, p-19.\\
11. NATO-Computer incident response capability (GNS):Cyber Newsletter, No-29, p-7, August 10, 2012.

\subsection{Interview and Comments}
1. German Newspaper, Die Zeit-Zeit Online:\\
 How secure is quantum cryptography, really?  \\
August 17, 2012. 
http://www.zeit.de/digital/internet/ 2012-08/quantenkryptographie\\
H.P.Yuen, O.Hirota, and R.Renner.\\
\\
2. German Newspaper, Die Zeit-Zeit Online:\\
 Physicist versus Cryptographers,  \\
August 17, 2012. 
http://www.zeit.de/digital/internet/ 2012-08/quantenkryptographie/seite-2\\
H.P.Yuen, O.Hirota\\
\\
3. Cryptographer (Associate professor in Brazil)\\
It seems to me that they found  evidence that QKD bits are not uniformly distributed, invalidating 
unconditional security.\\

4. Cryptographer (Professor in Germany)\\
So far, the definition of the security was unsubstantial, but now clear.


\end{document}